\documentclass[twocolumn]{aastex631}

\received{Aug 8, 2024}
\revised{Oct 2, 2024}
\accepted{Oct 17, 2024}

\begin{document}
\title{The radiative-convective gap: fact or fiction?}

\author[0000-0003-1939-6351]{Wolfgang Brandner}
\affiliation{Max-Planck-Institut f\"ur Astronomie, K\"onigstuhl 17, 69117 Heidelberg, Germany}
\author{Artur Sorg}
\affiliation{Englisches Institut, Rheinstra{\ss}e 14, 69126 Heidelberg, Germany}
\author{Siegfried R\"oser}
\affiliation{Zentrum für Astronomie der Universität Heidelberg, Astronomisches Rechen-Institut, Mönchhofstra{\ss}e 12-14, 69120 Heidelberg, Germany}
\author{Elena Schilbach}
\affiliation{Zentrum für Astronomie der Universität Heidelberg, Astronomisches Rechen-Institut, Mönchhofstra{\ss}e 12-14, 69120 Heidelberg, Germany}

\correspondingauthor{Wolfgang Brandner}
\email{brandner@mpia.de}

\begin{abstract}{
Gaia characterizes the stellar populations of nearby open clusters with unprecedented precision.
We investigate the B\"ohm-Vitense gap, which has been found as a prominent feature in the stellar sequence of open clusters.
Using PARSEC isochrone fitting, we derive astrophysical parameters for more than 1100 stars in Praesepe, identify more than 1100 bona fide single stars in the $\alpha$\,Persei (Melotte 20) open cluster, and confirm their approximate match in terms of age ($\approx$710\,Myr and $\approx$45\,Myr) and metallicity ([M/H] $\approx$+0.15\,dex and $\approx$+0.13\,dex) to the Hyades and Pleiades, respectively. By merging data of the cluster pairs, we improve number statistics. 
We do not find a clear gap in the combined observational G$_{\rm abs}$ vs.\ BP-RP color-magnitude diagram (CMD) in the stellar mass range corresponding to the location of the B\"ohm-Vitense gap. We reproduce gaps in simulated Hyades-type CMDs randomly drawn from an initial mass function. 
There is no strong evidence for a discontinuity originating in the transition from radiative to convective energy transport in the stellar photosphere. We conclude that the observed gaps in the stellar sequences of open clusters could be explained by small number statistics and the uneven mass-color relation at the transition from spectral type A to F.}
\end{abstract}


\section{Introduction} \label{sec:intro}
\cite{Stroemgren1956} and \cite{Mendoza1956} were among the first to comment on a discontinuity in the colors of nearby main-sequence field stars for 0.22$\lessapprox$ B-V $\lessapprox$ 0.31\,mag, i.e.\ late A- to early F-stars. 
By employing cumulative distribution functions in B-V \cite{BoehmVitense1974} identified reduced stellar incidences and gaps in the stellar sequences of the Hyades, Praesepe, $\alpha$\,Persei (Melotte 20), and Coma Berenices open clusters in the same approximate color range. \cite{Mazzei1988} found that gaps in stellar sequences could be the result of small population sizes and random sampling of the initial mass function. 

Using Hipparcos data \cite{Newberg1998} investigated magnitude and distance limited samples of main sequence field stars, and did not find any gaps in the color range B-V = 0.22 to 0.31\,mag.
\cite{DeBruijne2000} identified two gaps in a color-magnitude diagram (CMD) of a Hipparcos derived single star sample of the Hyades \citep{Perryman1998}, which they associated with the onset of surface convection and a sudden increase in the depths of the convection zone. 
\cite{Rachford2000} investigated whether or not small number statistics could explain the appearance of gaps. They concluded that while for individual clusters gaps at random colors could be due to stochastic, there was a very low probability for clear gaps to occur in the same color range by chance in six out of nine of the open clusters in their sample.

In the present paper we aim to investigate the authenticity of the gap aided by Gaia DR3's unprecedented astrometric and photometric quality. The structure of the paper is as follows. In section 2 we derive the single star sequences of the Praesepe and $\alpha$\,Persei open clusters, and determine their ensemble age and metallicity. In section 3 we investigate the merged sample of stars in the Hyades and Praesepe, and in the Pleiades and $\alpha$\,Persei open clusters, and compare with synthetic data sets. In section 4, we discuss the findings.


\section{Praesepe and $\alpha$\,Persei single star sequence} \label{sec:psss}

\begin{figure*}[ht!]
\epsscale{0.60}
\hbox{
\plotone{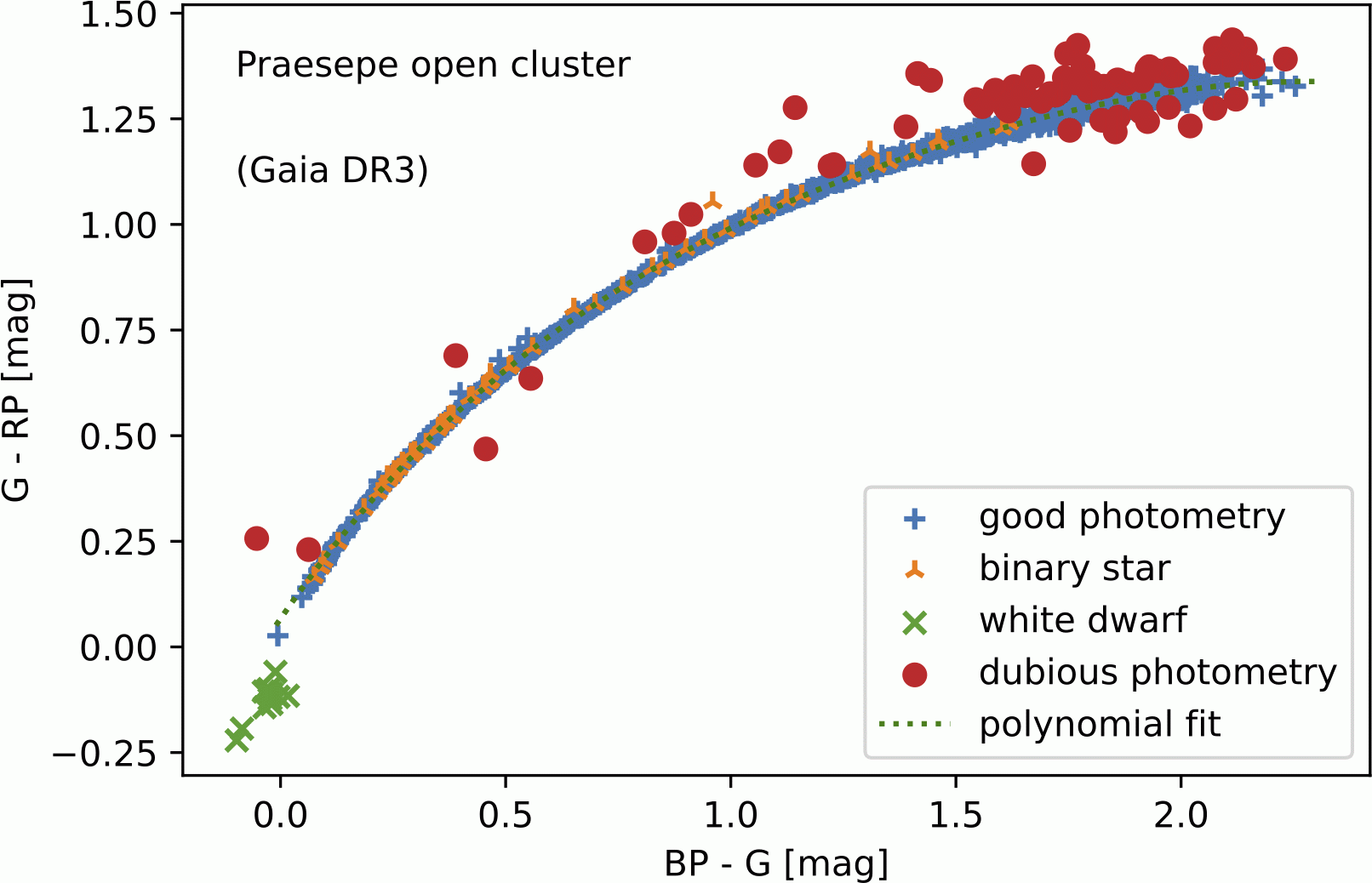}
\plotone{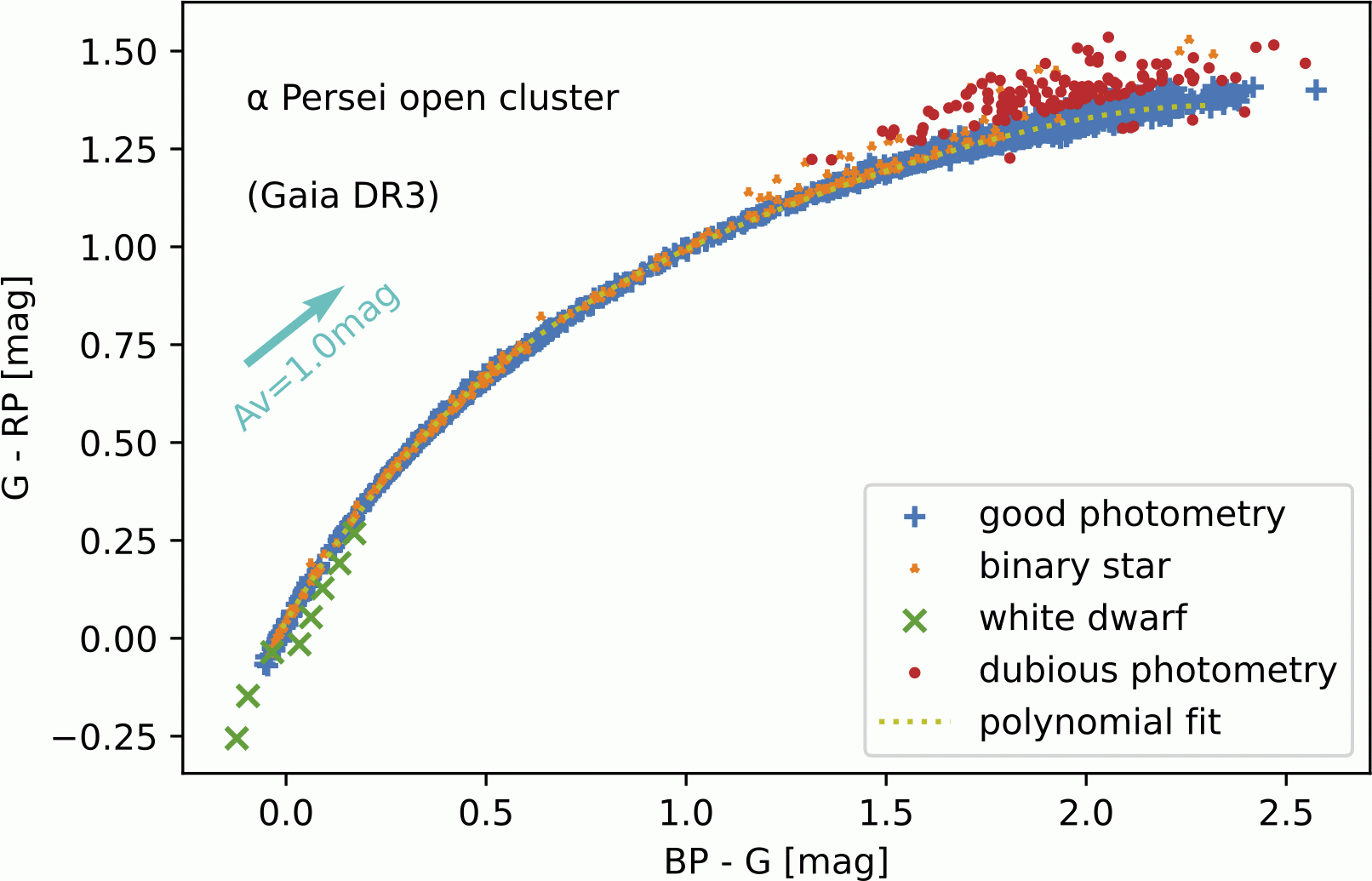}
}
\caption{Color-color diagrams of the Praesepe (left) and $\alpha$\,Persei (right) open clusters based on Gaia DR3 photometry. Blue plus and green crosses mark stars and white dwarfs, respectively, with valid photometry in all three Gaia photometric bands. Orange ``tri-ups'' indicate likely binary stars. Red circles mark stars with suspect photometry in at least one of the photometric bands. For $\alpha$\,Persei (A$_{\rm V}$=0.3\,mag) we plot an extinction vector corresponding to A$_{\rm V}$=1.0\,mag.
\label{fig:PraeTCD}}
\end{figure*}

\begin{table*}[htb]
\caption{Candidate members of the Praesepe cluster, classified according to {\it GAIA} DR3, sorted by RA.}             
\label{Prae_flags}      
\centering                          
\begin{tabular}{r c c c c c c c c c c}        
GAIA DR3 ID&  dpgeo & lo$\_$dpgeo & hi$\_$dpgeo & G & $\sigma_{\rm G}$& BP & $\sigma_{\rm BP}$ & RP & $\sigma_{\rm RP}$ & flag \\
            & [pc]     & [pc]      & [pc]       & [mag]&[mag]& [mag]&[mag]&[mag]&[mag]& \\
\hline                                   
504271599124398464& 215.006& 213.015& 217.050& 16.3920& 0.0007& 17.7144& 0.0079& 15.2687& 0.0023&1\\
655146893925330816& 213.231& 211.104& 215.943& 16.4955& 0.0007& 17.8141& 0.0081& 15.3661& 0.0019&2\\
1091089819680643200& 175.971& 175.534& 176.364& 12.7843& 0.0003& 13.3685& 0.0012& 12.0651& 0.0008&1\\
681114502418610688& 175.288& 170.792& 180.419& 17.6768& 0.0014& 19.5195& 0.0573& 16.3953& 0.0050&1\\
908834566101511552& 173.982& 171.967& 176.291& 16.5334& 0.0009& 18.1318& 0.0137& 15.3180& 0.0031&1\\
681144777643959296& 196.682& 195.440& 198.055& 15.4392& 0.0006& 16.6279& 0.0047& 14.3602& 0.0012&1\\
\end{tabular}
    \begin{quote}
    Median , low (lo, 16th quantile) and high (hi, 84th quantile) of the photo-geometric distance posterior dpgeo are from \cite{Bailer2021}; flag: 1 - bona fide single, 2 - likely binary or multiple, 3 - white dwarf, 4 - peculiar {\it GAIA} DR3 BP-G vs.\ G-RP color
      \end{quote}
\tablecomments{This table is published in its entirety in the machine-readable format. The first six entries are shown here for guidance regarding its form and content.}
\end{table*}

\begin{table*}[htb]
\caption{Candidate members of the $\alpha$\,Persei cluster, classified according to {\it GAIA} DR3, sorted by RA.}             
\label{AlfPer_flags}      
\centering                          
\begin{tabular}{r c c c c c c c c c c}        
GAIA DR3 ID&  dpgeo & lo$\_$dpgeo & hi$\_$dpgeo & G & $\sigma_{\rm G}$& BP & $\sigma_{\rm BP}$ & RP & $\sigma_{\rm RP}$ & flag \\
            & [pc]     & [pc]      & [pc]       & [mag]&[mag]& [mag]&[mag]&[mag]&[mag]& \\
\hline                                   
406045284749433856& 176.748& 171.585& 180.866& 18.3730& 0.0018& 20.4029& 0.0969& 17.0370& 0.0076& 1\\
407807355212400640& 165.063& 164.438& 165.733& 9.8411& 0.0004& 10.1004& 0.0011& 9.4217& 0.0007& 1\\
407807389572138368& 165.301& 164.879& 165.706& 10.7702& 0.0008& 11.1051& 0.0026& 10.2705& 0.0020& 1\\
357148761414557952& 168.853& 167.407& 170.121& 16.3567& 0.0008& 17.7840& 0.0094& 15.1840& 0.0030& 2\\
349857594869311232& 177.012& 173.994& 180.985& 17.2351& 0.0013& 19.0660& 0.0434& 15.9323& 0.0055& 1\\
357721164592501888& 168.960& 166.460& 171.534& 17.1857& 0.0013& 19.0376& 0.0240& 15.8994& 0.0042& 1\\
\end{tabular}
    \begin{quote}
    Median , low (lo, 16th quantile) and high (hi, 84th quantile) of the photo-geometric distance posterior dpgeo are from \cite{Bailer2021}; flag: 1 - bona fide single, 2 - likely binary or multiple, 3 - white dwarf, 4 - peculiar {\it GAIA} DR3 BP-G vs.\ G-RP color
      \end{quote}
\tablecomments{This table is published in its entirety in the machine-readable format. The first six entries are shown here for guidance regarding its form and content.}
\end{table*}

Our sample comprises 1393 likely members of the Praesepe cluster and its tidal tails defined by \cite{Roeser2019} and 1573 likely members of the $\alpha$\,Persei cluster identified by us using the same methodology. We process the sample following the steps devised for the Hyades \citep{Brandner2023a,Brandner2023b} and Pleiades \citep{Brandner2023c} clusters. 

i) In a 1st step, we flag as likely astrometric binaries all sources with DR3 Renormalized Unit Weight Error (RUWE, \cite{Lindegren2021}) larger than 1.4. Among the $\alpha$\,Persei astrometric candidate members we also flag 124 stars with discrepant DR3 radial velocities as potential (spectroscopic) binaries. In addition we flag sources with BP-G$<$0.1\,mag and G-RP$<$0.0\,mag in Praesepe and sources with G$>$17\,mag and BP-RP$<$0.45\,mag in $\alpha$\,Persei as likely white dwarfs.

ii) Next we fit a 4th order polynomial to the stellar sequence in BP-G vs.\ G-RP color-color space, and apply iterative $\sigma$-clipping to identify and flag outliers with suspect photometry (Figure \ref{fig:PraeTCD}).

iii) In a 3rd step, we calculate the absolute magnitude for all stars using the photo-geometric distance estimates by \cite{Bailer2021}.

iv) The next step is similar to one of the methods discussed by \cite{Valle2021}. We fit an 8th order polynomial to the stellar sequence in BP-RP vs.\ G$_{\rm abs}$ color-absolute magnitude space and apply again an iterative sigma-clipping.  Stars rejected by the $\sigma$-clipping are flagged as potential photometric binary stars. For Praesepe this results in 1150 stars flagged as bona fide single stars, 160 stars flagged as potential astrometric or photometric binaries, 13 stars flagged as likely white dwarfs, and 70 stars flagged as having suspect photometry in at least one of the DR3 G, BP, and RP photometric bands (Table \ref{Prae_flags}). In $\alpha$\,Persei we identify  1105 bona fide single stars, 339 potential astrometric, RV, or photometric binaries, eight white dwarfs, and 121 stars with suspect photometry (Table \ref{AlfPer_flags}).

\begin{figure*}[ht!]
\plotone{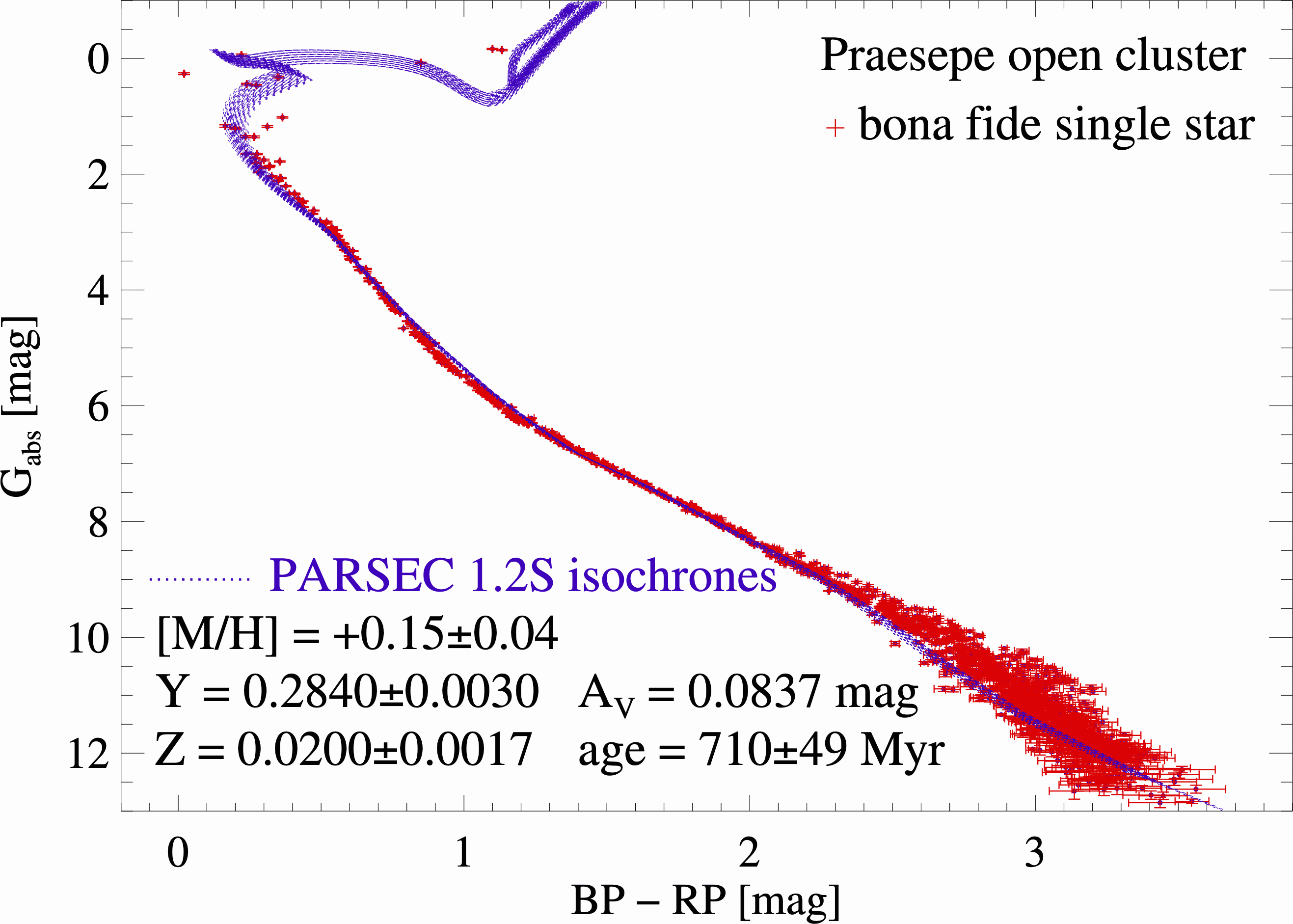}
\caption{Absolute color-magnitude diagram of the bona fide single stars in the Praesepe open cluster based on Gaia DR3. The uncertainties in absolute magnitude and color are indicated by red error bars. Overplotted as dotted blue lines is the family of best fitting PARSEC isochrones.
\label{fig:PraeCMDiso}}
\end{figure*}
\begin{figure}[ht!]
\plotone{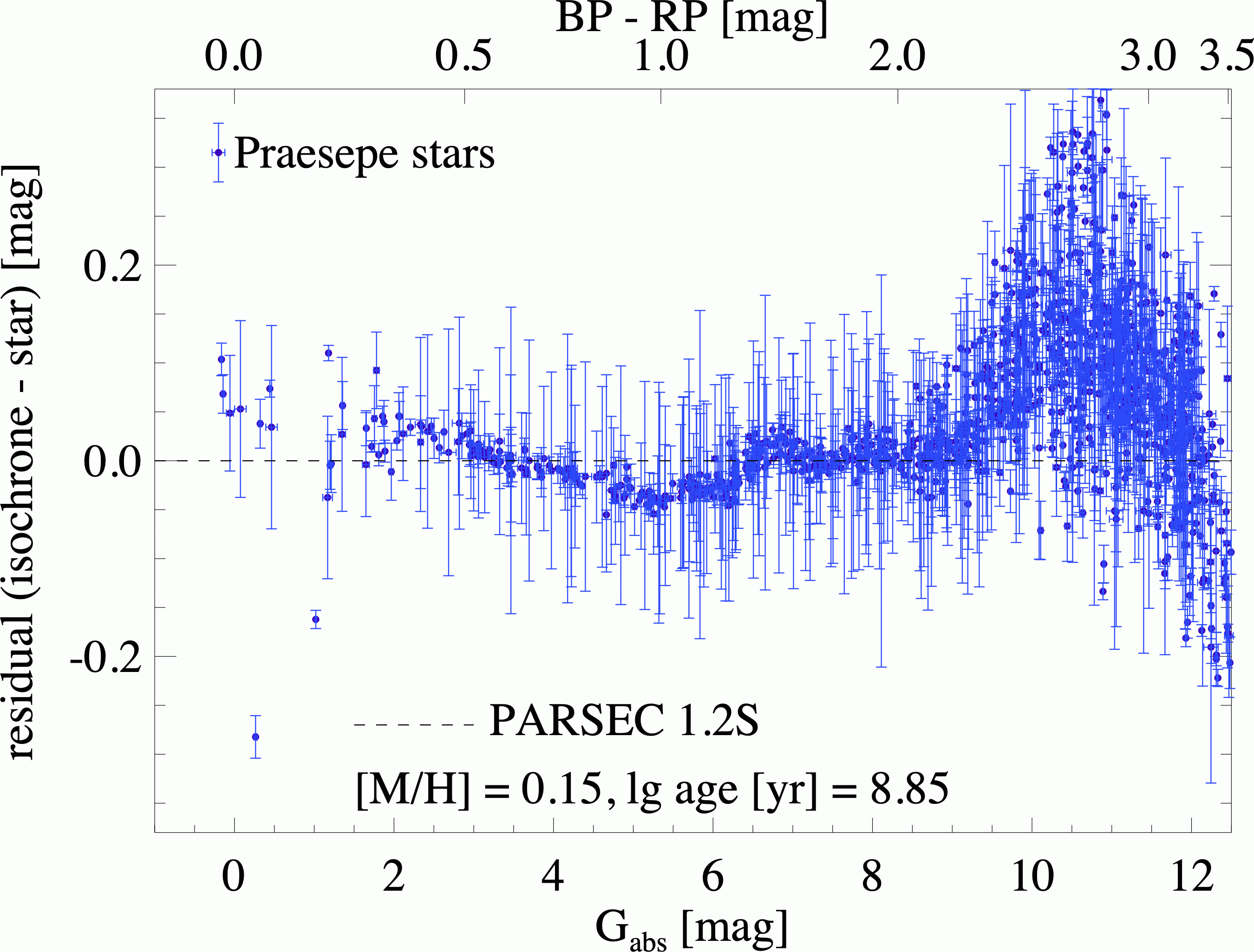}
\caption{The minimum distance in color-magnitude space of stars on the single star sequence in Praesepe from the best fitting PARSEC isochrone.
\label{fig:PraeCMDres}}
\end{figure}


v) As a final step we identify the family of best-fitting PAdova and TRieste Stellar Evolution Code (PARSEC) 1.2S isochrones \citep{Giradi2002,Marigo2008,Bressan2012} to determine ensemble age and metallicity of the Praesepe and $\alpha$\,Persei open clusters. Table \ref{Extinction} summarizes the interstellar extinction values assumed for each of the clusters when computing the PARSEC isochrones. $\alpha$\,Persei stands out as having the highest extinction, which also varies across the face of the cluster \citep{Prosser1992}.
Due to the accelerated stellar evolution towards higher masses, most of the information on the age of the stellar population of Praesepe is encoded in the more massive stars. 
Therefore we select a subsample of bona fide single stars with G$_{\rm abs} \le$5.5\,mag, and use $\chi ^2$ minimization in color-absolute magnitude space to identify the family of best-fitting non-rotating PARSEC isochrones. This yields [M/H] = +0.15$\pm$0.04 and lg\,age\,[Myr] = $8.85\pm 0.03$ for the Praesepe cluster (Figure \ref{fig:PraeCMDiso}). In Figure \ref{fig:PraeCMDres} we show the minimum distance in color-magnitude space of stars on the single star sequence in Praesepe from the best
fitting PARSEC isochrone. In the color range BP-RP = 0.3 to 0.7 mag, the maximum deviation is $\lessapprox$0.04 mag. For $\alpha$\,Persei isochrone fitting yields [M/H] = +0.13$\pm$0.03 and lg\,age\,[Myr] = $7.63\pm 0.07$.

\begin{table*}[htb]
\caption{Cluster bulk interstellar extinction}             
\label{Extinction}      
\centering                          
\begin{tabular}{l l l l l l}        
cluster&  A$_{\rm BP}$ & A$_{\rm G}$ &A$_{\rm RP}$ & A$_{\rm V}$ & A$_{\rm V}$ reference \\
            & [mag]     & [mag]      & [mag]       & [mag]& \\
\hline  
$\alpha$\,Persei &0.325 &0.251 &0.190 &0.30 & \cite{Prosser1992,Lodieu2019}\\
Hyades &0.0034 &0.0026 &0.0020 &0.0031 &\cite{Taylor2006} \\
Pleiades &0.13 &0.10 &0.076 &0.12 &\cite{Meynet1993,vanLeeuwen2009} \\
Praesepe &0.0907 &0.0700 & 0.0531 &0.0837 &\cite{Taylor2006} \\
\end{tabular}
    \begin{quote}
    Assumed Gaia A$_\lambda$/A$_{\rm V}$ ratios are the same as used for the reddening of the PARSEC isochrones.
      \end{quote}
\end{table*}


In Tables \ref{tab:Prae_metal_age} and \ref{tab:AlfPer_metal_age} we compare the age and abundance values with literature estimates derived from isochrone fitting. We also list spectroscopic abundance estimates based on solar-type dwarfs. While for Praesepe the majority of the abundance determinations achieve statistical uncertainties in the range of 0.01 to 0.04\,dex, the divergence in the weighted mean estimates points at the presence of larger systematic uncertainties between the different studies. Overall stellar interior and atmospheric models seem to provide good relative consistency when applied to homogeneous data sets as is the case with our analyses of Gaia DR3 data sets of nearby open clusters using PARSEC 1.2S isochrones. Our new values for metallicity and age of Praesepe and $\alpha$\,Persei are in overall agreement with previous studies. Almost all estimates suggest a slightly super-solar metallicity, and an age in the range 600 to 750\,Myr for Praesepe, and 20 to 80\,Myr for $\alpha$\,Persei. 

The results for individual stellar masses, effective temperatures, luminosities, and surface gravities for stars in Praesepe are summarized in Table \ref{tab:Prae_single}. In order to avoid ambiguities in the evolutionary status of main-sequence turn-off and post-main sequence stars, we restrict this analysis to 1129 bona fide single stars with $1.0 \le$G$_{\rm abs} \le$12.5\,mag.

\begin{table*}
\caption{Compilation of abundance, $\alpha_{\rm ML}$, and age estimates for the Praesepe cluster. In the top we summarize estimates based on isochrone fitting. In the lower part of the table we list spectroscopic abundance estimates.}             
\label{tab:Prae_metal_age}      
\centering                          
\begin{tabular}{l l c c c l}        
[Fe/H] & $\alpha_{\rm ML}$ & mass range & age  &PD$^1$&reference \\
 &     & [M$_\odot$] & [Myr] & & \\ \hline
+0.03              &1.64  &[0.67,2.60] &$757 \pm 36$ &SD&\cite{Gaspard2009}\\
+0.08          &1.9  &[0.75,2.60]  &$665${\raisebox{0.5ex}{\tiny$^{+14}_{-7}$}} &Jo&\cite{Bell2014}\\
$+0.30 \pm 0.06$&  1.82 &[1.75,2.40]  &603$\pm$42 &TY&\cite{Gossage2018}\\
$+0.09 \pm 0.01$&  1.82&[1.25,2.40]  &679$\pm$60 &2M&\cite{Gossage2018}\\
$+0.15 \pm 0.04$$^2$  &1.74 &[0.90,2.45]&710$\pm$49&G3&this paper\\ \hline
$+0.11 \pm 0.01^3$  & &[0.97,1.05]&---&---&\cite{An2007}\\
$+0.27 \pm 0.10$  & &[1.02,1.14]&---&---&\cite{Pace2008}\\
$+0.12 \pm 0.03^3$  & &[0.98,1.10]&---&---&\cite{Boesgaard2013}\\
$+0.15 \pm 0.02^3$  & &[0.73,1.08]&---&---&\cite{Gebran2019}\\
$+0.21 \pm 0.02^3$  & &[0.94,1.18]&---&---&\cite{DOrazi2020}\\
$+0.19 \pm 0.02^3$  & &[1.00,1.18]&---&---&\cite{Vejar2021}\\ \hline
\end{tabular}
    \begin{quote}
        $^1$ key to photometric data set (PD): 2M - based on 2MASS photometry \citep{Cutri2003}; G3 - based on Gaia DR3 \citep{GAIA2016,GAIA2023} ; Jo - based on {\it (U)BV} photometry \citep{Johnson1952}; TY - based on {\it TYCHO} photometry \citep{Hog2000}; $^2$ default PARSEC 1.2S solar scaled composition Y=0.2485+1.78Z with Z$_\odot$=0.0152 \citep{Bressan2012}, and assuming [Fe/H] = [M/H], i.e.\ no offset. $^3$ weighted mean abundance
      \end{quote}
\end{table*}

\begin{table*}
\caption{Compilation of abundance, $\alpha_{\rm ML}$, and age estimates for the $\alpha$\,Persei cluster. In the top we summarize estimates based on isochrone fitting. In the lower part of the table we list spectroscopic abundance estimates.}             
\label{tab:AlfPer_metal_age}      
\centering                          
\begin{tabular}{l l c c c l}        
[Fe/H] & $\alpha_{\rm ML}$ & mass range & age  &PD$^1$&reference \\
 &     & [M$_\odot$] & [Myr] & & \\ \hline
& & &51$\pm$6 &UBV&\cite{Mermilliod1981}\\
& &[0.7,7.0] &52 &UBV&\cite{Meynet1993}\\
& & &71$\pm$11 &UBV&\cite{Loktin2001}\\
 & &[0.9,7.0]  &20 &TY&\cite{Sanner2001}\\
 & &  &35 &ASCC&\cite{Kharchenko2005}\\
$+0.06$ & &[0.9,7.0]  &52 &TY&\cite{Makarov2006}\\
$+0.14$ & &[0.9,7.0]  &71 &G2&\cite{Gaia2018}\\
$+0.13 \pm 0.03$$^2$  &1.74 &[0.9,7.0]&43$\pm$7&G3&this paper\\ \hline
$-0.054 \pm 0.046^3$  & &[1.2,1.7]&---&---&\cite{Boesgaard1990}\\ 
$+0.07 \pm 0.09$  & & &---&---&\cite{Fu2022}\\
$+0.01 \pm 0.05$  & &&---&---&\cite{Netopil2022}\\ \hline
\end{tabular}
    \begin{quote}
        $^1$ key to photometric data set (PD): ASCC - based on TYCHO and ground based photometry \citep{Kharchenko2001};  G2/3 - based on Gaia DR2/3 \citep{GAIA2016,GAIA2023};  TY - based on {\it TYCHO} photometry \citep{Hog2000}; UBV - groud based UBV photometry (various sources); $^2$ default PARSEC 1.2S solar scaled composition Y=0.2485+1.78Z with Z$_\odot$=0.0152 \citep{Bressan2012}, and assuming [Fe/H] = [M/H], i.e.\ no offset. $^3$ weighted mean abundance
      \end{quote}
\end{table*}

\begin{deluxetable*}{rlllhhrhhhhhrrrrrrrr}
\tabletypesize{\scriptsize}
\tablewidth{0pt} 
\tablecaption{Astrophysical parameters of bona fide single stars in the Praesepe open cluster \label{tab:Prae_single}}
\tablehead{
\colhead{GAIA DR3 ID} & \colhead{RA}& \colhead{DEC} & \colhead{dpgeo} & \nocolhead{e_dpgeo} & \nocolhead{E\_d} & \colhead{G} & \nocolhead{e\_G} & \nocolhead{BP} & \nocolhead{e\_BP} & \nocolhead{RP} & \nocolhead{e\_RP} & \colhead{mass} & \colhead{$\sigma_{\rm mass}$}  & \colhead{log T$_{\rm eff}$} & \colhead{$\sigma_{\rm logTeff}$} & \colhead{log L} & \colhead{$\sigma_{\rm log L}$} & \colhead{log g} & \colhead{$\sigma_{\rm log g}$} \\
\colhead{} & \colhead{(deg)} & \colhead{(deg)} & \colhead{(pc)} & \nocolhead{(pc)} & \nocolhead{(pc)}& \colhead{(mag)} & \nocolhead{(mag)} & \nocolhead{(mag)}& \nocolhead{(mag)}& \nocolhead{(mag)}& \nocolhead{(mag)} & \colhead{(M$_\odot$)}  & \colhead{(M$_\odot$)} & \colhead{(K)}  & \colhead{(K)} & \colhead{(L$_\odot$)}  & \colhead{(L$_\odot$)} & \colhead{(cm/s$^2$)}  & \colhead{(cm/s$^2$)}\\
} 
\startdata 
504271599124398464&  86.307076&  77.055432& 215.006& 213.015& 217.050&  16.3920&  0.0007&  17.7144&  0.0079&  15.2687&  0.0023&  0.4670&  0.0169&  3.5308&  0.0044& -1.6150&  0.0494&  4.7966&  0.0174\\
655146893925330816& 121.092236&  16.007482& 213.231& 211.104& 215.943&  16.4955&  0.0007&  17.8141&  0.0081&  15.3661&  0.0019&  0.4545&  0.0204&  3.5276&  0.0051& -1.6521&  0.0600&  4.8093&  0.0212\\
1091089819680643200& 121.800698&  62.673647& 175.971& 175.534& 176.364&  12.7843&  0.0003&  13.3685&  0.0012&  12.0651&  0.0008&  0.7823&  0.0061&  3.6744&  0.0021& -0.6597&  0.0121&  4.6409&  0.0018\\
681114502418610688& 122.188684&  25.303456& 175.288& 170.792& 180.419&  17.6768&  0.0014&  19.5195&  0.0573&  16.3953&  0.0050&  0.2737&  0.0386&  3.4870&  0.0102& -2.1899&  0.1253&  4.9652&  0.0219\\
908834566101511552& 122.452402&  38.649424& 173.982& 171.967& 176.291&  16.5334&  0.0009&  18.1318&  0.0137&  15.3180&  0.0031&  0.3934&  0.0231&  3.5137&  0.0049& -1.8313&  0.0662&  4.8710&  0.0228\\
681144777643959296& 122.512470&  25.545449& 196.682& 195.440& 198.055&  15.4392&  0.0006&  16.6279&  0.0047&  14.3602&  0.0012&  0.5433&  0.0106&  3.5552&  0.0038& -1.3796&  0.0317&  4.7244&  0.0088\\
\enddata
\tablecomments{This table is published in its entirety in the machine-readable format. The first six entries are shown here for guidance regarding its form and content. In the displayed table we omit the columns with uncertainties on the distance estimates, and part of the Gaia DR3 photometry.}
\end{deluxetable*}

\section{Combined Praesepe-Hyades and $\alpha$\,Persei-Pleiades data}

\begin{figure*}[ht!]
\epsscale{0.60}
\hbox{
\plotone{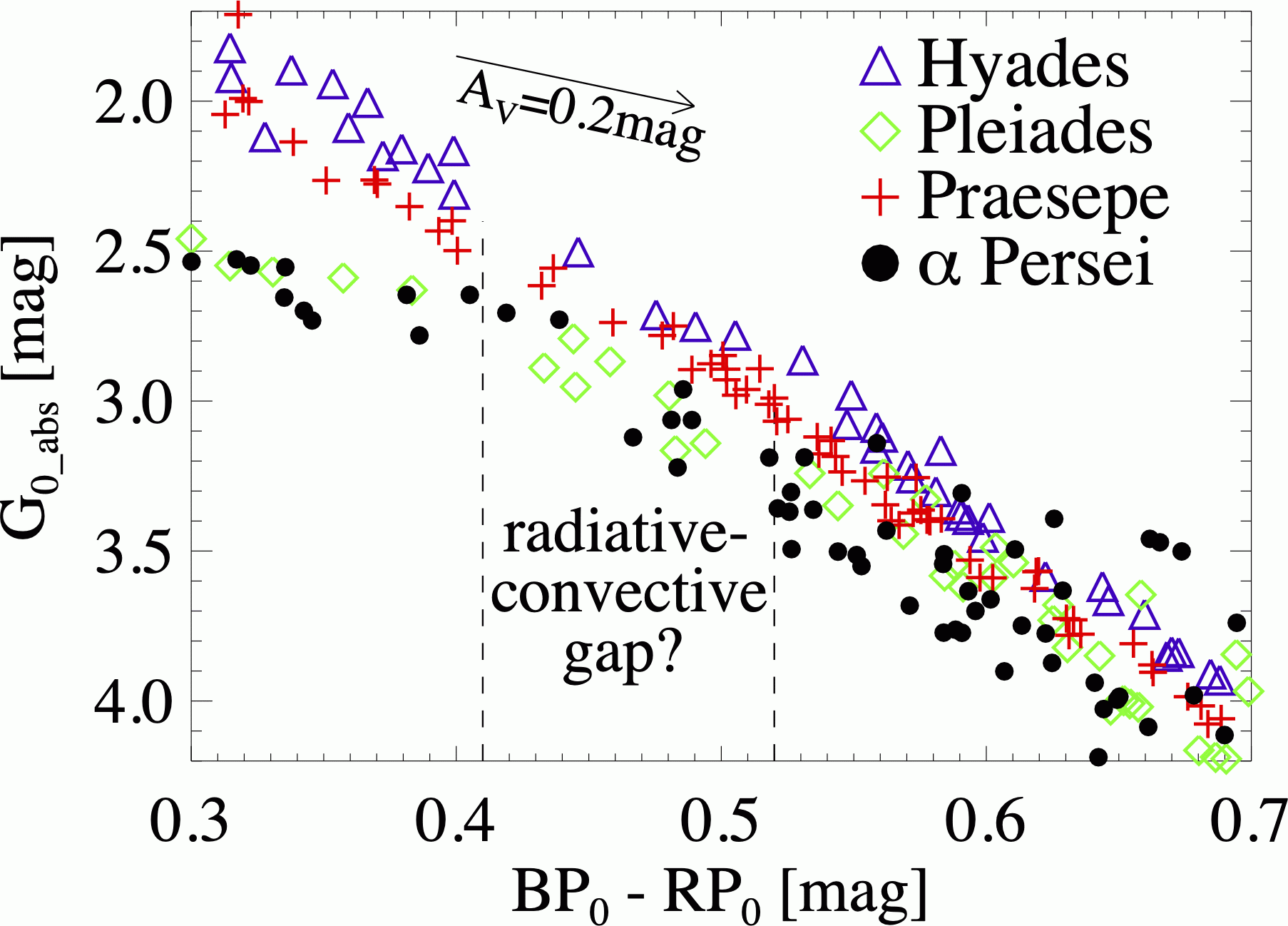}
\plotone{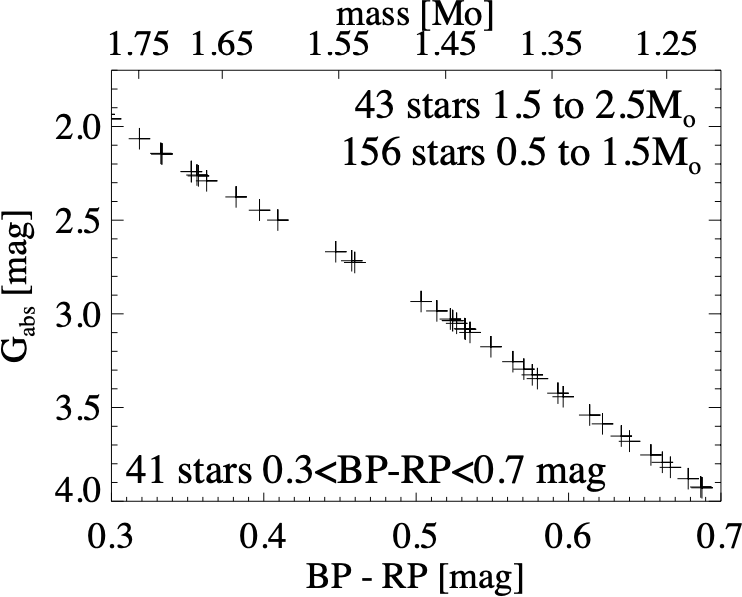}
}
\caption{Left: Color-magnitude diagram of the Hyades, Praesepe, Pleiades, and $\alpha$\,Persei clusters for $0.3\le$BP$_0$-RP$_0 \le 0.7$\,mag. Stars in Praesepe fill-in the gap in the Hyades stellar sequence. Similarly, the combined Praesepe and $\alpha$\,Persei sequence does not exhibit a pronounced gap.} Right: One instance of Monte Carlo simulations of clusters as populous as the Hyades in the mass range 0.5 to 2.5\,M$_\odot$. The stochastic population of the stellar sequence according to the initial mass function results in a gap similar to the gap in the Hyades.
\label{fig:RadConGap}
\end{figure*}

The close match in age and metallicity of the Hyades and Praesepe open clusters (see also \cite{Brandt2015}) allows us to merge both data sets. Figure \ref{fig:RadConGap}, left, shows the combined CMD for both clusters, zoomed in on the color range $0.3\le$BP$_0$-RP$_0$$\le 0.7$\,mag. In this color range, we find 40 and 63 bona fide singles stars in the Hyades and Praesepe, respectively. For comparison, we also include  37 stars in the Pleiades cluster \citep{Brandner2023c} and 58 stars in the $\alpha$\,Persei cluster.
The gap in the Hyades sequence between BP-RP $\approx$ 0.41 and 0.52\,mag is filled in by stars in Praesepe. A pronounced gap is also absent in the combined sequence of the younger $\alpha$\,Persei and Pleiades clusters. Given the similarity between the Hyades and Praesepe, this raises the question if the Hyades gap(s) might be the result of statistical fluctuations and low number statistics.

To investigate this we carried out Monte Carlo simulations emulating Hyades-type clusters.
We assume an Initial Mass Function (IMF) with $\Gamma = -1.3$ ($\alpha = 2.3$) in the mass range 0.5 to 2.5\,M$_\odot$ \citep{Salpeter1955,Kroupa2002}. The lower mass limit of 0.5\,M$_\odot$ is still close to the 100\% completeness level for the Hyades, while 2.5\,M$_\odot$ defines the main-sequence turn-off. We count 199 bona fide single stars in this mass range in the Hyades. The simulated cluster sequences result from random drawing of the IMF. We derive color and absolute brightness of each stars by linear interpolation of the grid points of the best-fitting PARSEC isochrone. In the mass range of interest, the grid sampling is $\Delta$mass = 0.02 to 0.03\,M$_\odot$. 

\begin{figure*}[ht!]
\plotone{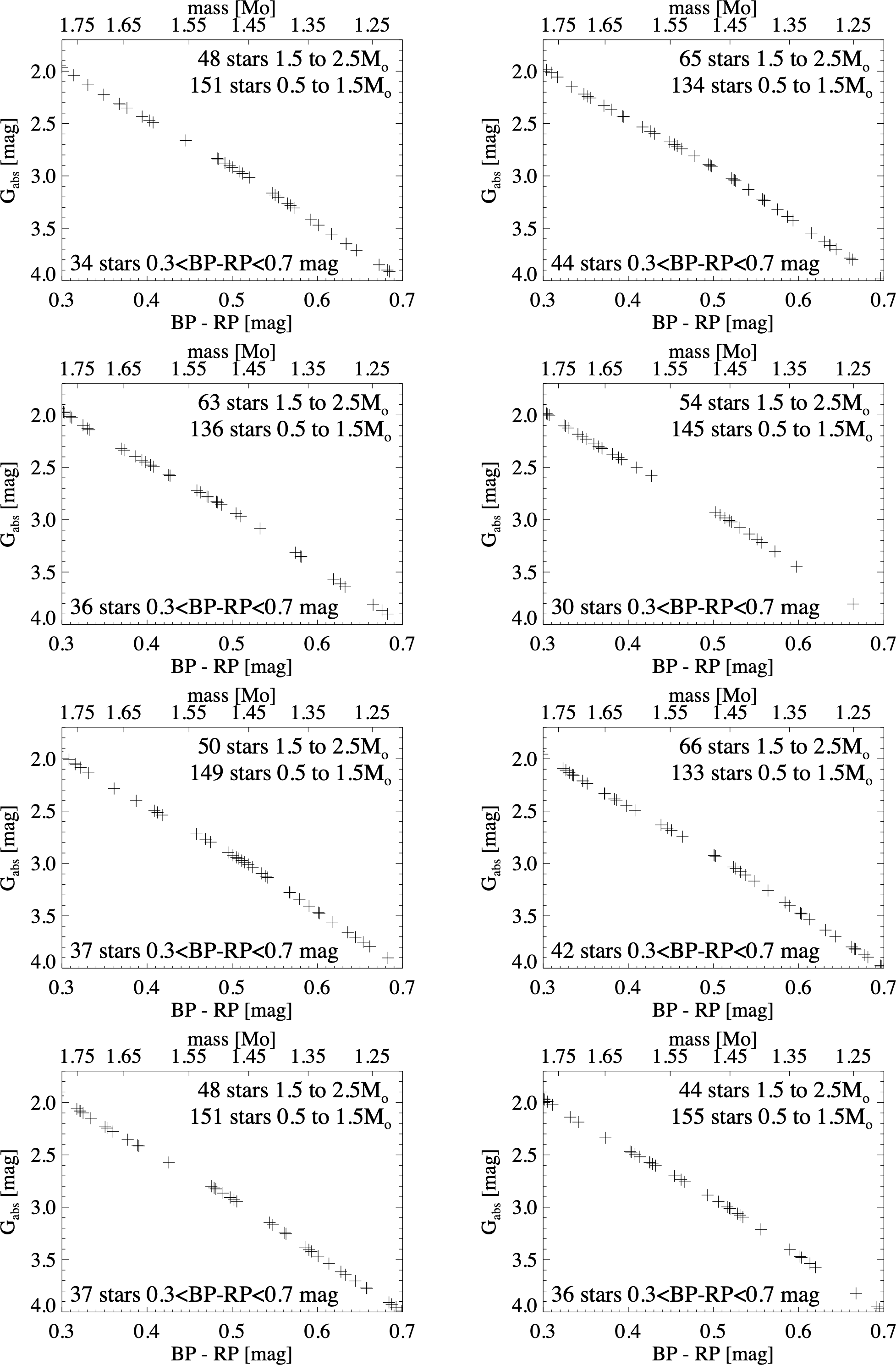}
\caption{Instances of Monte Carlo-generated synthetic color-magnitude diagrams of a Hyades-type cluster. The random drawing from an initial mass function combined with small number statistics, and the uneven relation between stellar mass and BP-RP color can result in gaps in the sequence. 
\label{fig:SimCMD}}
\end{figure*}

A scan by eye reveals gaps in the color range BP-RP = 0.4 to 0.5\,mag in about 40\% to 45\% of 128 simulated CMDs, while another 40\% to 45\% of simulated CMDs show gaps only outside of this color range. Continuous stellar sequences without any gaps in the color range BP-RP = 0.3 to 0.7\,mag are found in less than 10\% of the simulations. Figure \ref{fig:RadConGap}, right, showcases one instance of the simulations, which resulted in a gap in the CMD similar to the one observed in the Hyades. In Fig.\ \ref{fig:SimCMD} we show eight randomly selected instances of the resulting CMDs. 
An example for a smooth stellar sequence without any major gaps is presented in the simulation on the right, 1st from top. Other examples show relatively wide gaps in the BP-RP color range 0.4 to 0.5\,mag. 

\begin{figure*}[ht!]
\epsscale{0.60}
\hbox{
\plotone{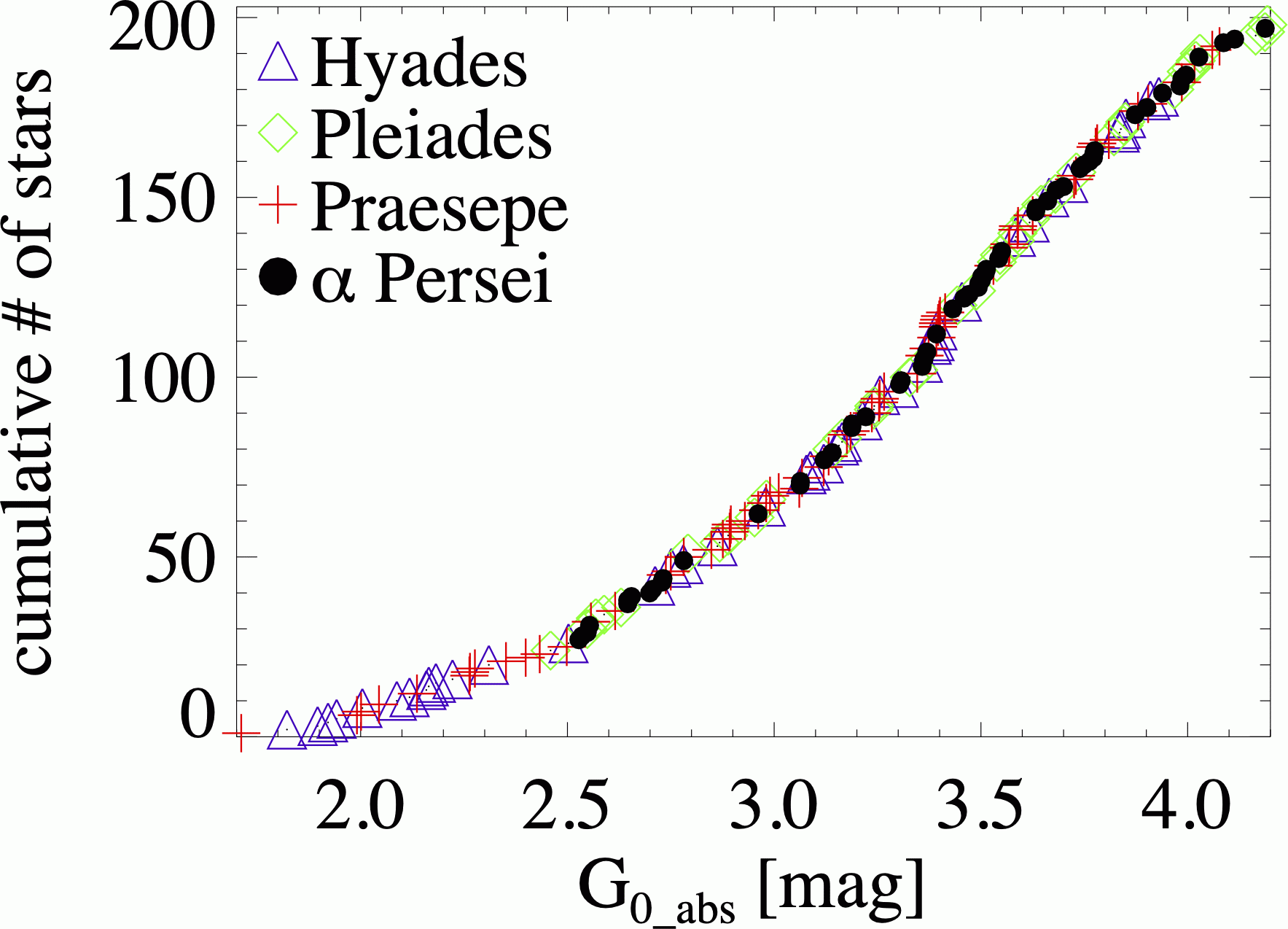}
\plotone{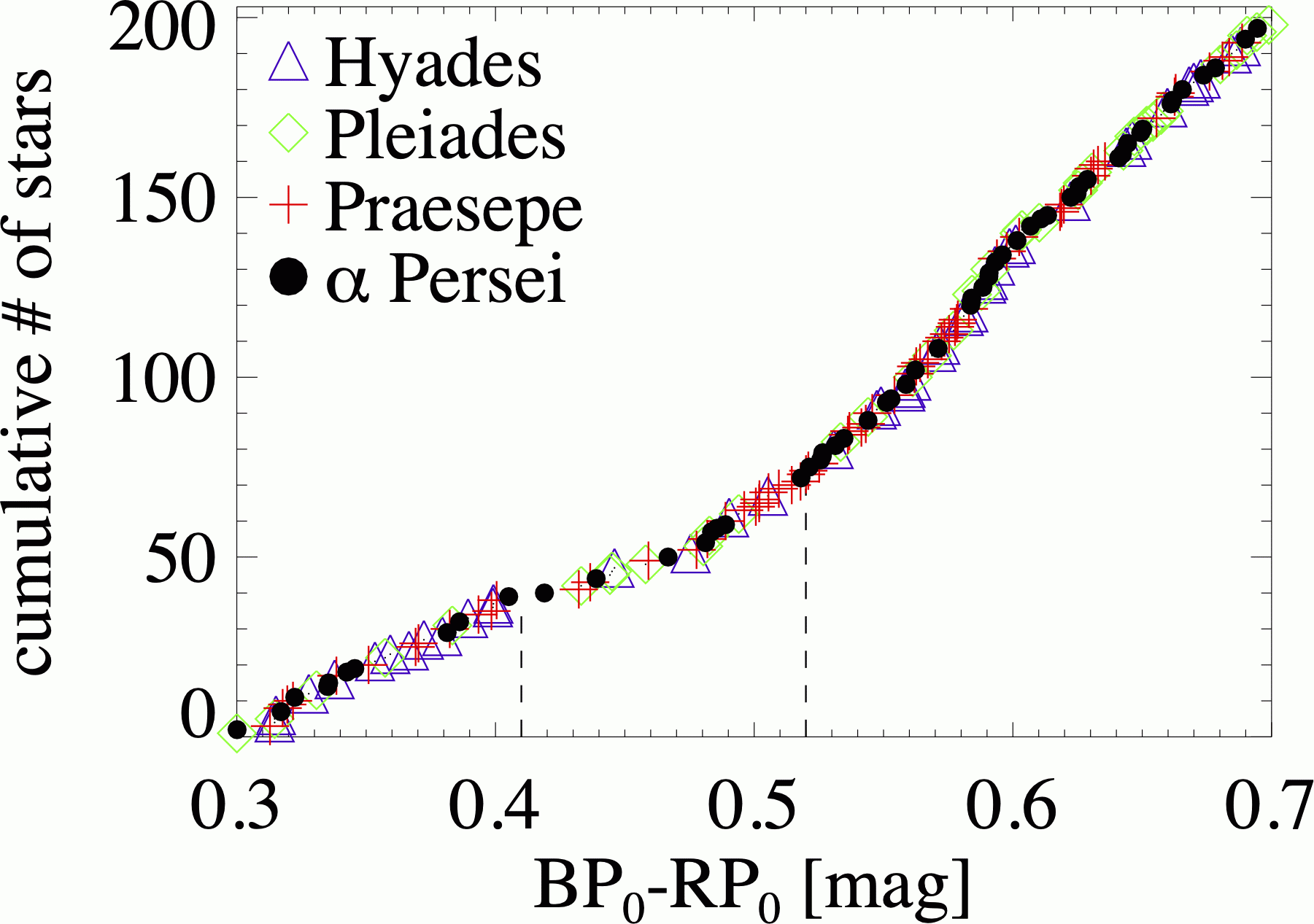}
}
\caption{Cumulative distribution of extinction corrected absolute G$_{\rm abs}$ magnitudes (left) and BP-RP colors (right) of bona fide single stars in the Hyades, Pleiades, Praesepe, and $\alpha$\,Persei. 
The vertical dashed lines mark the color range of the ``radiative-convective'' gap seen in the Hyades CMD. 
\label{fig:CumulativeDiagrams}}
\end{figure*}

\section{Discussion} \label{sec:discussion}

It has long been argued that the gaps seen in color-magnitude space in the stellar sequences of the Hyades and other open clusters might have a physical origin tracing back to transitions in the dominant mode of energy transportation in the outer layers of stars. 

Using atmospheric models based on mixing length theory (MLT) \cite{BoehmVitense1958,BoehmVitense1970} identified an abrupt transition from radiation to convection dominated outer layers, and an associated jump in B-V color from $\approx$0.20 to 0.28\,mag, with decreasing effective temperature for early F-type main-sequence stars at T$_{\rm eff} \approx$7500$\pm$500\,K. \cite{DAntona2002} simulated the effect of convection in deeper layers of the stellar envelope using full spectrum of turbulence (FST) models \citep{Canuto1996}. In Monte Carlo simulations with 100 stars each drawn at random from a Salpeter IMF in the mass range 1.2 to 1.7\,M$_\odot$, they identified a significant ``T$_{\rm eff}$ gap'' around $\approx$6650\,K (B-V$\approx$0.35 to 0.40\,mag), which coincides with the second (redder) gap found by \cite{DeBruijne2000}. They note, though, that such a gap did not show up in their MLT-based simulations. 

The non-detection of a gap in MLT-based simulations by \cite{DAntona2002} suggests that the (non-) occurrence of gaps depends on the precise mapping of stellar mass to effective temperature and color, and hence is very sensitive to the underlying atmospheric models. 
Common to these studies is that they are based on 1D LTE models, which approximate convection, assume plane parallel atmospheres, and also, e.g., do not take into account the change in effective surface gravity with latitude in fast rotating stars. Inherent to 1D models is that they tend to be more stable, and to show sharper transitions than 3D models \citep[see, e.~g.][]{Trampedach2014,Rodriguez2024}. 

The gaps are preferentially observed at the transition between spectral types A and F, which also marks the transition from neutral H to H$^-$ becoming the dominant continuum opacity source. Furthermore, spectral type F is marked by the emergence of spectral lines of Ca\,II, Mn\,I, and Fe\,I, which all contribute to lowering the flux in particular in the blue (B and BP), thus resulting in a more ``pronounced'' transition to redder BP-RP colors from A- to F-spectral class. The resulting uneven relation between stellar mass and BP-RP color is apparent when one compares the top (mass) and bottom (color) abscissae of the simulated CMDs in Figure \ref{fig:RadConGap}, right, or \ref{fig:SimCMD}. The step from 1.55 to 1.65\,M$_\odot$ corresponds to a much larger $\Delta$(BP-RP) than the step from 1.65 to 1.75\,M$_\odot$, resulting in an reduced probability of stars with BP-RP$\approx$0.45\,mag compared to stars with BP-RP$\approx$0.35\,mag. In Figure \ref{fig:CumulativeDiagrams} the combined extinction corrected stellar cluster data are presented as cumulative distributions of absolute brightness (left) and color (right). The distribution of brightness values reveals a steepening at G$_{\rm 0\_abs}$=2.5\,mag, which is explained by the lack of stars with BP$_0$-RP$_0 > 0.3$\,mag, which are brighter than G$_{\rm 0\_abs}=$2.5\,mag in the younger $\alpha$\,Persei and Pleiades clusters. 

The cumulative color distribution reveals a shallower slope in the color range BP$_0$-RP$_0 \approx$0.40\,mag to 0.48\,mag, which naturally results from the IMF-to-color relation. This combined with the small number statistic could explain the gaps found in nearby open clusters. Our Monte Carlo simulations seem to confirm this. 

Our analysis suggests that there is a continuous transition in B-V and BP-RP color rather than a discontinuity, and that it traces back to the change in the dominant continuum opacity source and the emergence of certain spectral lines at the transition from A to F. The emergence of a convective outer layer might be regarded as a symptom rather than the cause. As pointed out by \cite{Jermyn2022}, there is also a need to revisit convection with state-of-the-art 3D stellar models.

\begin{acknowledgments}
AS acknowledges support by the Hector Seminar.
This work has made use of data from the European Space Agency (ESA) mission
{\it Gaia} (\url{https://www.cosmos.esa.int/gaia}), processed by the {\it Gaia}
Data Processing and Analysis Consortium (DPAC,
\url{https://www.cosmos.esa.int/web/gaia/dpac/consortium}). Funding for the DPAC
has been provided by national institutions, in particular the institutions
participating in the {\it Gaia} Multilateral Agreement.
\end{acknowledgments}

\vspace{5mm}
\facilities{Gaia}

\bibliography{lit}{}
\bibliographystyle{aasjournal}

\end{document}